\documentclass[superscriptaddress]{revtex4}
\usepackage[utf8]{inputenc}
\usepackage{amsmath}
\usepackage{mathtools}
\usepackage{braket}
\usepackage{graphicx}
\usepackage{pgfplots}
\usepackage{csquotes}
\usepackage{hhline}
\usepackage{amssymb}

\usepackage{dcolumn}
\usepackage{tabularx}
\usepackage[colorlinks=true,linkcolor=red,citecolor=magenta,urlcolor=blue]{hyperref}
\usepackage{longtable}
\usepackage{braket}
\usepackage{float}
\usepackage{avm}

\def\benum#1\eenum{\begin{enumerate}#1\end{enumerate}}
\def\be#1\ee{\begin{equation}#1\end{equation}}
\def\bml#1\eml{\begin{multline}#1\end{multline}}
\def\ba#1\ea{\begin{align}#1\end{align}}
\def\bas#1\eas{\begin{align*}#1\end{align*}}

\def\bit{\begin{itemize}}
\def\eit{\end{itemize}}

\begin{document}

\title{Discrimination of Highly Entangled Z-states in IBM Quantum Computer}
\author{Saipriya Satyajit}
\affiliation{Department of Physics, Indian Institute of Technology Bombay, Mumbai 400076, India\\}
  \author{Karthik Srinivasan}
  \affiliation{Department of Physics, Indian Institute of Technology Madras, Chennai 600036, India}
    \author{Bikash K. Behera}
    \affiliation{Department of Physical Sciences, Indian Institute of Science Education and Research Kolkata,\\ Mohanpur 741246, India}
    \author{Prasanta K. Panigrahi}
  \affiliation{Department of Physical Sciences, Indian Institute of Science Education and Research Kolkata,\\ Mohanpur 741246, India}

\begin{abstract}
Measurement-based quantum computation (MQC) is a leading paradigm for building a quantum computer. Cluster states being used in this context act as one-way quantum computers. Here, we consider Z-states as a type of highly entangled states like cluster states, which can be used for one-way or measurement based quantum computation. We define Z-state basis as a set of orthonormal states which are as equally entangled as the cluster states. We design new quantum circuits to non-destructively discriminate these highly entangled Z-states. The proposed quantum circuits can be generalized for N-qubit quantum system. We confirm the preservation of Z-states after the performance of the circuit by quantum state tomography process.
\end{abstract}


\maketitle
\section{Introduction}
Entangled states have a wide range of application in the field of quantum computation and quantum information \cite{NCbook}. Using highly entangled states like Bell states, GHZ states, cluster states and Brown \emph{et al.} states quantum information processing tasks such as quantum teleportation \cite{BENNETPRL1993,GhoshNJP2002,SMPRA2008,SCJPAMT2009, SMEPJD2011, PAULQIP2011}, quantum secret sharing \cite{SMPRA2008,SCJPAMT2009, SMEPJD2011, JainEPL2009, PRASATHQP2012}, quantum information splitting \cite{SMPRA2008,SMEPJD2011, PAULQIP2011,PKPPJP2009,SMOC2010}, super dense coding \cite{SMPRA2008, PAULQIP2011, AGRAWALPRA2006}, quantum cheque \cite{SRMQP2016, BKBQIP2017} etc. have been performed. Recently, a set of schemes discriminating orthogonal entangled states \cite{GuptaIJQI2007,PKPACP2006,MunroJOBQSO2005,WangQIP2013,ZhengIJTP2016} have been proposed.  

Non-destructive discrimination of orthogonal entangled states is a significant variant of discrimination of orthogonal entangled states. Using this protocol, we can discriminate orthogonal entangled states without disturbing them using indirect measurements on ancillary qubits, which contain information about the entangled states. This protocol plays a pivotal role in quantum information processing and quantum computation \cite{NCbook}. This has been proposed for generalized orthonormal qudit Bell state discrimination \cite{PKPACP2006} and has also been experimentally achieved for 2-qubit Bell states using both NMR \cite{GuptaIJQI2007,SamalIOP2010} and five qubit IBM quantum computer \cite{SisodiaPLA2017}. Some of the schemes have also been realized in optical medium \cite{LiJPBAMOP2000} by using Kerr type nonlinearity. 

This Distributed measurement technique finds a number of applications in photonic systems, measurement-based quantum computation, quantum error correction, Bell state discrimination across a quantum network involving multiple parties and optimization of the quantum communication complexity for performing measurements in distributed quantum computing. Non-destructive discrimination of entangled states has found applications in secure quantum conversation as proposed by Jain \emph{et al.} \cite{JainEPL2009}, which involves two communicating parties. A more general scenario involving multi-parties has also been demonstrated by Luo \emph{et al.} \cite{LuoQIP2014}.

On the other hand, cluster states also have many impressive applications, mainly in areas like measurement-based quantum computation (MQC) \cite{NielsenPLA2003,LeungIJQI2004,RoussendorfQIC2002,RoussendorfPRA2003,AliferisPRA2004,DanosJACM2007,ChildsPRA2005,BrowneNJP2007,VerstraetePRA2004,GrossPRL2007} or one-way quantum computation. MQC provides a promising conceptual framework which can face the theoretical and experimental challenges for developing useful and practical quantum computers that can perform daily-life computational tasks and solve real world problems. The exciting feature of MQC is that it allows quantum information processing by cascaded measurements on qubits stored in a highly entangled state. The scheme of a one-way quantum computer was first introduced by Raussendorf and Briegel \cite{RaussendorfPRL2001}, whose key physical resource was the cluster state \cite{BriegelPRL2001}. The scheme used measurements which were central elements to perform quantum computation \cite{NielsenPRL1997,GottesmanNAT1999,KnillNAT2001}. In the proposed scheme, the cluster state can be used only once as it does not preserve entanglement after one-qubit measurements, hence the name one-way quantum computation.       

Researchers are currently using IBM Quantum Experience to perform various quantum computational and quantum informational tasks \cite{RundlePRA2017,Solano2arXiv2017,GrimaldiSD2001,KarlaarXiv2017,GosharXiv2017,GangopadhyayarXiv2017,SchuldarXiv2017,MajumderarXiv2017,LiarXiv2017,SisodiaarXiv201794,VishnuarXiv2017,GedikarXiv2017,BKB6arXiv2017,BKB7arXiv2017,BKB8arXiv2017,SolanoQMQM2017}. Test of Leggett-Garg \cite{HuffmanPRA2017} and Mermin inequality \cite{AlsinaPRA2016}, non-Abelian braiding of surface code defects \cite{WoottonQST2017}, entropic uncertainty and measurement reversibility \cite{BertaNJP2016} and entanglement assisted invariance \cite{DeffnerHel2017} have been illustrated. Estimation of molecular ground state energy \cite{KandalaNAT2017} and error correction with 15 qubit repetition code \cite{WoottonarXiv2017} have also been implemented using 16 qubit IBM quantum computer ibmqx5. Experimental realization of discrimination of Bell states has motivated us to define a new set of highly entangled orthogonal states as Z-states and to demonstrate their discrimination using IBM's five-qubit real quantum processor \emph{ibmqx4}. In the present work, we define Z-states which also form an orthonormal basis for the corresponding N-qubit quantum system. We then propose new quantum circuits to distinguish between those orthogonal states and experimentally verify our theoretical results using ibmqx4. We also generalize the quantum circuit for discrimnating N-qubit Z-states.

The rest of the paper is organized as follows. In Sec. \ref{sec2}, we define Z-states and design quantum circuits for creating these states. In Sec. \ref{sec3}, we propose a new quantum circuit for non-destructively discriminating Z-states. Following which, we explicate the experimental process by taking a particular Z-state and verify the results by quantum state tomography. Finally, we conclude our paper by discussing some future applications of Z-states which can be realized in a real quantum computer.

\section{Z- STATES}
The expression for N-qubit cluster state is given below. 

\begin{equation}
\label{1}
\ket{C_N}=  \frac{1}{\sqrt{2^N}} \ \otimes_{a=1} ^{N} (\ket{0}_{a} Z_{a+1} + \ket{1}_{a}) 
\end{equation}

2-qubit, 3-qubit and 4-qubit cluster states are expressed as,

\begin{equation}
 \ket{C_2} = \frac{1}{\sqrt{2}} \ (\ket{0}\ket{+}~+~\ket{1}\ket{-}) ,\linebreak
 \end{equation}
 \begin{equation}
 \ket{C_3} = \frac{1}{\sqrt{2}} \ (\ket{+}\ket{0}\ket{+} ~+~ \ket{-}\ket{1}\ket{-}) ,
 \end{equation}
 
 \begin{equation}\label{4}
 \begin{aligned}
 \ket{C_4} = \frac{1}{\sqrt{2}} \ (\ket{+}\ket{0}\ket{+}\ket{0} ~+~ \ket{+}\ket{0}\ket{-}\ket{1} ~+~ \\ \ket{-}\ket{1}\ket{-}\ket{0} ~+~ \ket{-}\ket{1}\ket{+}\ket{1}) 
 \end{aligned}
 \end{equation}

\begin{figure}[H]
\centering
\includegraphics[scale=0.65]{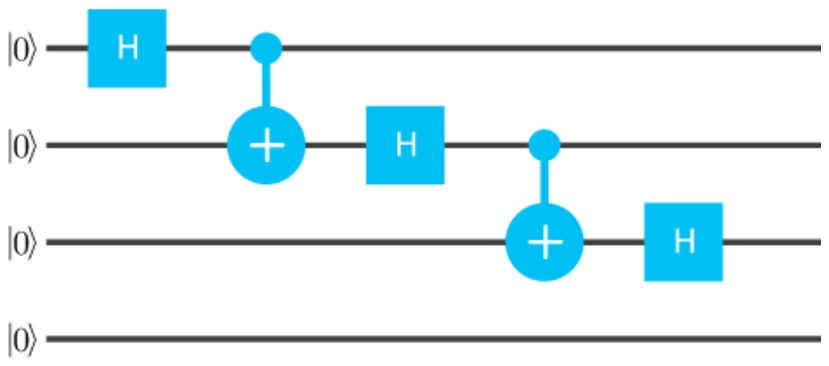}
\caption{\emph{Circuit depicting the creation of \emph{$C_3$}~ State}}
\label{Figure1}
\end{figure}

A circuit creating 3-qubit cluster state is shown in Fig. \ref{Figure1}.  

For any given number of qubits, the cluster state is not the only highly entangled state. By applying an appropriate number of Z gates (phase flip gates) on different qubits we can generate Z-states. These states are as entangled as their corresponding cluster state. The Z-states are in an equal superposition of all the computational basis states, which form an orthonormal basis for the Hilbert space. 
The 2 qubit Z-states in the computational basis are as follows.  

\begin{equation}
\ket{Z_{2}^0} = [1,1,1,-1]^T 
\end{equation}
\begin{equation}
\ket{Z_{2}^1} = [1,1,-1,1]^T   
\end{equation}
\begin{equation}
\ket{Z_{2}^2} = [1,-1,1,1]^T
\end{equation}
\begin{equation}
\ket{Z_{2}^3} = [1,-1,-1,-1]^T
\end{equation}

The Z-states for three-qubit case are as shown below.
\begin{equation}
\ket{Z_{3}^0} = [1,1,1,-1,1,1,-1,1]^T 
\end{equation}
\begin{equation}
\ket{Z_{3}^1} = [1,-1,1,1,1,-1,-1,-1]^T   
\end{equation}
\begin{equation}
\ket{Z_{3}^2} = [1,1,1,-1,-1,-1,1,-1]^T
\end{equation}
\begin{equation}
\ket{Z_{3}^3} = [1,-1,1,1,-1,1,1,1]^T
\end{equation}
\begin{equation}
\ket{Z_{3}^4} = [1,1,-1,1,1,1,1,-1]^T 
\end{equation}
\begin{equation}
\ket{Z_{3}^5} = [1,-1,-1,-1,1,-1,1,1]^T   
\end{equation}
\begin{equation}
\ket{Z_{3}^6} = [1,1,-1,1,-1,-1,-1,1]^T
\end{equation}
\begin{equation}
\ket{Z_{3}^7} = [1,-1,-1,-1,-1,1,-1,-1]^T
\end{equation}

\begin{figure}[H]
\centering
\includegraphics[scale=0.6]{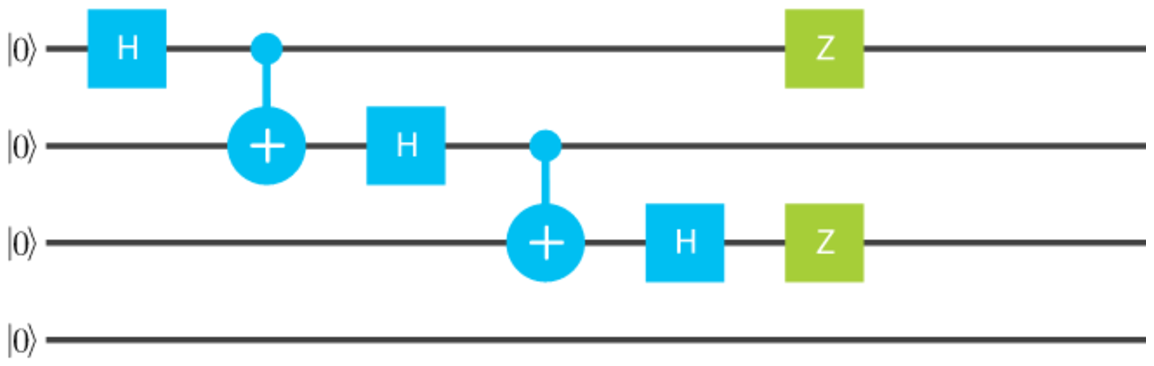}
\caption{\emph{Quantum circuit to create $\ket{Z_{3}^3}$ state.}}
\label{Figure2}
\end{figure}

As stated above the Z-states can be created by applying Z gates to the appropriate qubits. For example, the circuit generating a 3-qubit Z-state, $\ket{Z_{3}^3}$ is illustrated in Fig. \ref{Figure2}. 

\section{QUANTUM CIRCUITS AND METHOD USED FOR NONDESTRUCTIVE DISCRIMINATION OF Z-STATES}
\begin{figure}[H]
\centering
\includegraphics[scale=0.47]{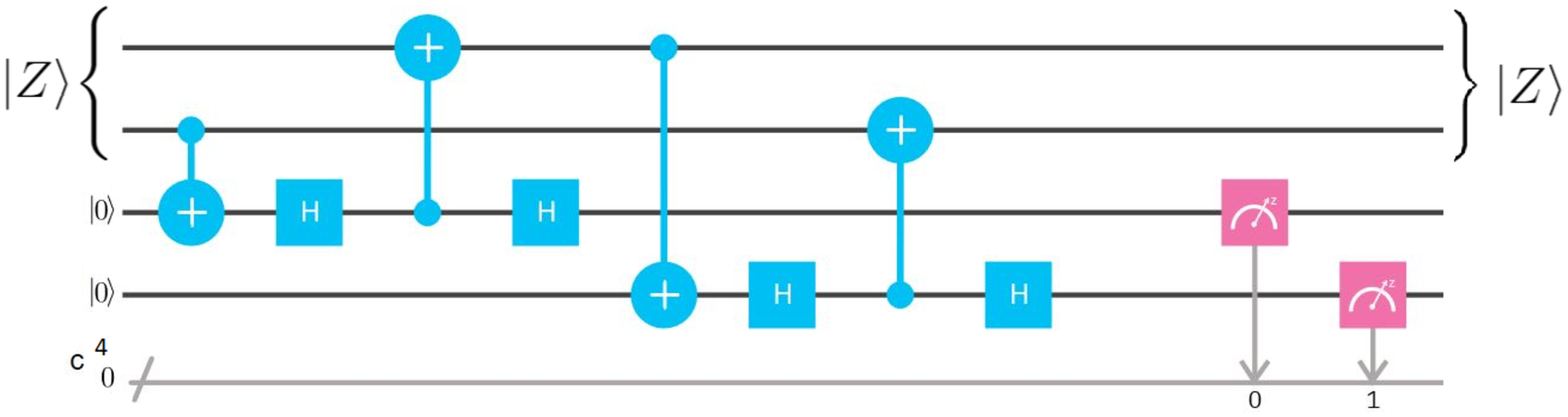}
\caption{\emph{A two qubit Z-state discrimination circuit}}
\label{Figure3}
\end{figure}

\begin{figure}[H]
\centering
\includegraphics[scale=0.6]{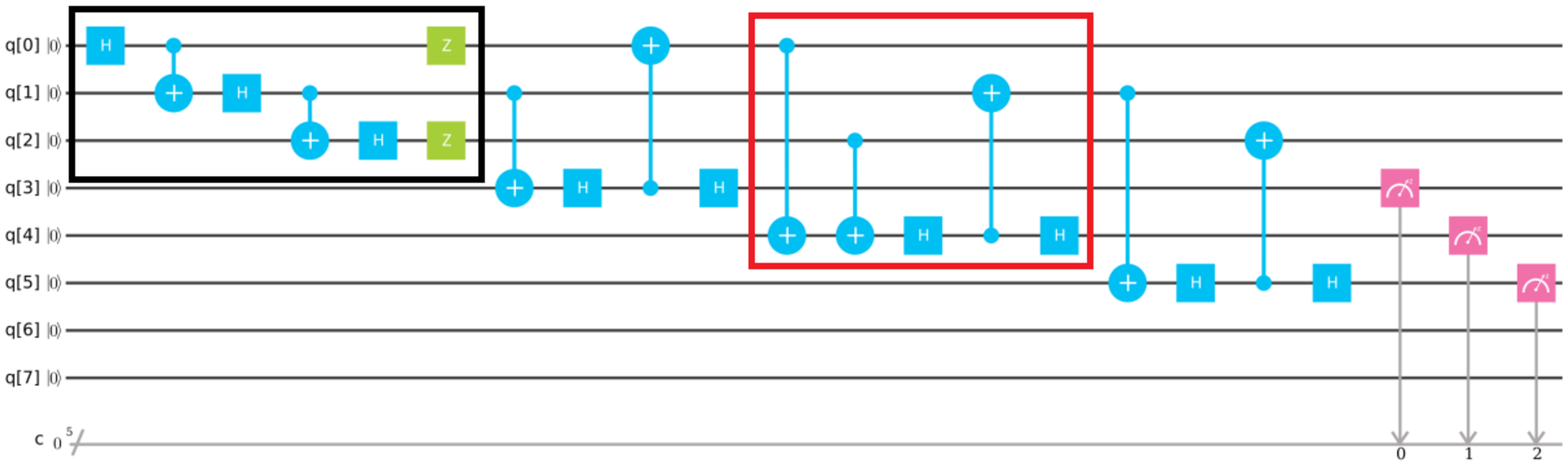}
\caption{\emph{A three qubit Z-state discrimination circuit}}
\label{Figure4}
\end{figure}

\begin{figure}[H]
\centering
\includegraphics[scale=0.5]{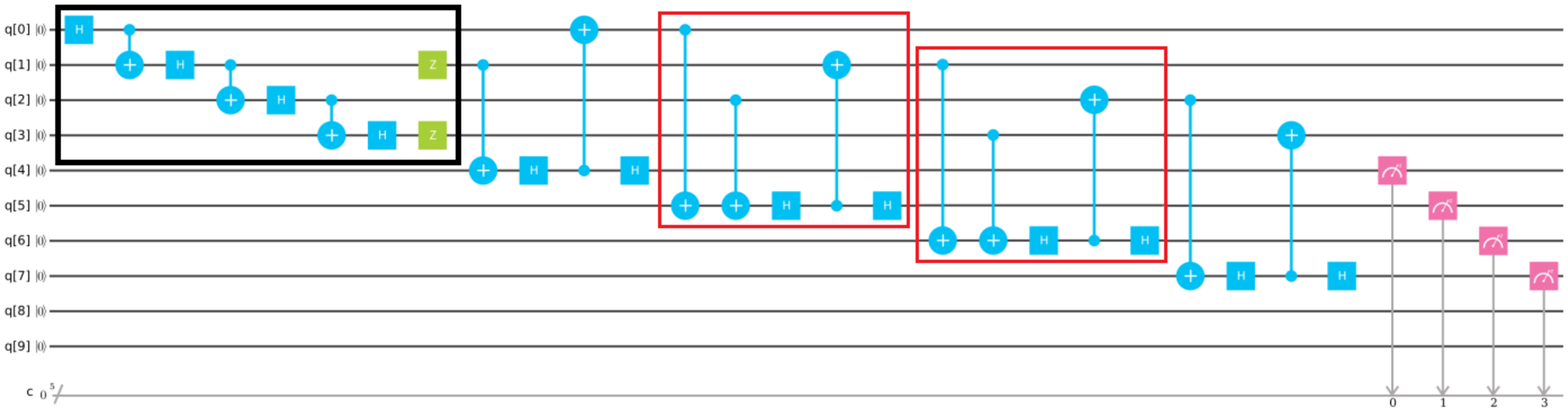}
\caption{\emph{A four qubit Z-state discrimination circuit}}
\label{Figure5}
\end{figure}

Since the Z-states are as equally entangled as the cluster state, we should be able to use those states for almost every quantum computational task which uses cluster states. Hence being able to discriminate between the Z-states non-destructively is important. The circuit illustrating discrmination of 2 qubit, 3 qubit and 4 qubit Z-states are shown in Figs. \ref{Figure3}, \ref{Figure4} and \ref{Figure5} respectively. The output of the circuit shown in Fig. \ref{Figure3} for each 2 qubit Z-state is shown in the Table 1.

\begin{table}[H]

\centering
\begin{tabular}{ |p{1cm}||p{4.4cm}|p{1.4cm}|}
 \hline
  S.No.&\hspace{1.7cm}Z-state& Ancilla\\
 \hline
  \hspace{0.3cm}1.&$\ket{00}+\ket{01}+\ket{10}-\ket{11}$&\hspace{0.4cm}00\\
  \hspace{0.3cm}2.&$\ket{00}+\ket{01}-\ket{10}+\ket{11}$&\hspace{0.4cm}01\\
  \hspace{0.3cm}3.&$\ket{00}-\ket{01}+\ket{10}+\ket{11}$&\hspace{0.4cm}10\\
  \hspace{0.3cm}4.&$\ket{00}-\ket{01}-\ket{10}-\ket{11}$&\hspace{0.4cm}11\\
 \hline
 \end{tabular}\\
 \caption{\emph{2 qubit Z-states and corresponding ancilla states}}
\label{table:1}
\end{table}

In Fig. \ref{Figure5}, the part of the circuit shown in black refers to the creation of one of the sixteen 4-qubit Z-states.
From the circuits discriminating 3 qubit and 4 qubit Z-states, we can see a pattern arising. It is clear that the methodology for getting information for the first and the last ancillary qubits remains the same. The pattern that emerges applies for rest of the ancillary qubits. We can see that for all the ancillary qubits except the first and the last, two CNOTs act on it with control qubits as qubits exactly one above and below the ancillary qubit's corresponding Z-state qubit. Then we apply a Hadamard on the ancillary qubit, CNOT on the corresponding qubit (which appears to be sandwiched between the controls of previous two CNOTs) with the ancillary qubit as control, followed by a Hadamard. This ``sandwich" pattern is enveloped in the red box. 

For discriminating between the Z-states of higher number of qubits we can simply keep repeating this sandwich pattern for the ancillary qubits other than the first and the last, while the circuit for the first and the last ancillary qubits remains the same.

\section{RESULTS}
\begin{figure}[H]
\centering
\includegraphics[scale=0.4]{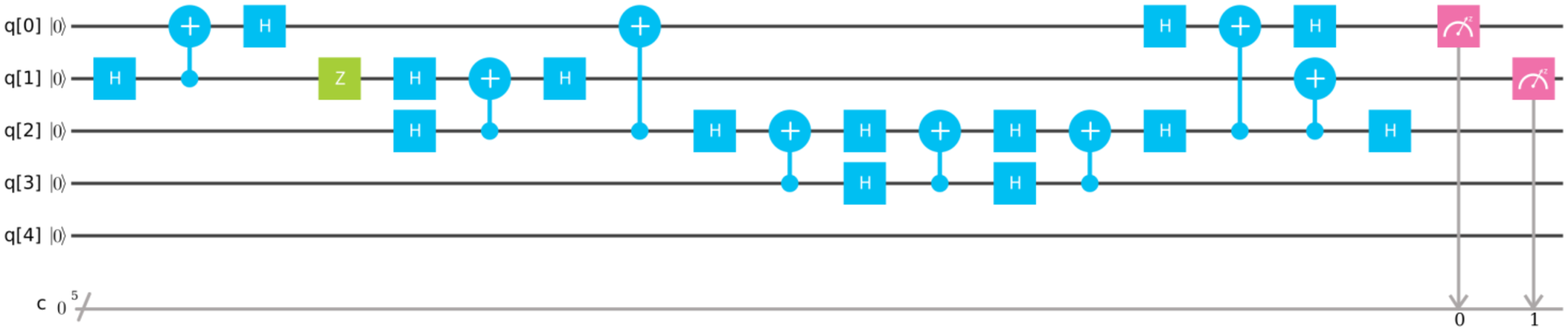}
\caption{Circuit designed in ibmqx4 to measure the 2-qubit Z-state }
\label{Figure6}
\end{figure}

We initially prepare the two qubit Z-state, $\ket{Z_2^1}$ with two ancillas in state $\ket{00}$. Then the necessary single qubit and two qubit quantum gates are applied on the Z-state and the ancillas. After the performance of the quantum circuit the above Z-state is measured to check the non destructiveness of the proposed protocol. For this purpose quantum state tomography is performed and it is observed that the Z-state remains undisturbed throughout the execution of the quantum circuit. Fig. \ref{Figure6} depicts the implementation of the proposed algorithm for non destructive discrimination of $\ket{Z_2^1}$ state using ibmqx4.  

The quantum circuit has been run and simulated 8192 times with different measurement basis and the following results have been obtained.
The theoretical ($\rho^T$) and experimental (simulational-$\rho_S^E$, and run-$\rho_R^E$)
density matrices are provided below.\\
\[\rho^T = \ket{Z_2^1}\bra{Z_2^1}\]

\[\begin{split}
\rho_{S}^{E} =
\left[{\begin{array}{cccc}
                   0.2460&-0.2495&0.2500&0.2515\\
                  -0.2495&0.2520&-0.2485&-0.2500\\
                  0.2500&-0.2485&0.2530&0.2505\\
                  0.2515&-0.2500&0.2505&0.2490\\
\end{array} } \right] +
\end{split}
\]
\[
\begin{split}
 \\ i\left[ {\begin{array}{cccc}
                              0.0000&0.0080&-0.0050&-0.0013\\
                              -0.0080&0.0000&-0.0033&-0.0030\\
                              0.0056&0.0033&0.0000&0.0000\\
                              0.0013&0.0030&0.0000&0.0000\\
\end{array} } \right]
\end{split}
\]

 \[
 \begin{split}
   \rho_{R}^{E}=
  \left[ {\begin{array}{cccc}
   0.3187 & -0.1015 & 0.1778 & 0.1520 \\
   -0.1015 & 0.2348 & -0.1250 & -0.1177
   \\
   0.1778 & -0.1250 & 0.2427 & 0.1545
   \\
   0.1520 & -0.1177 & 0.1545 & 0.2037\\
  \end{array} } \right]+
  \end{split}
\]
\[
 \begin{split}
  \\ i\left[ {\begin{array}{cccc}
   0.0000 & -0.0675 & 0.0000 & 0.0617\\
   0.0675 & 0.0000 & 0.0127 & -0.0150 \\
   0.0000 & -0.0127 & 0.0000 & 0.0155\\
   -0.0617 & 0.0150 & -0.0155 & 0.0000\\
  \end{array} } \right]
  \end{split}
 \]

The corresponding density matrices are plotted in the following Fig. \ref{Figure8}

\begin{figure}[H]
\centering
\includegraphics[scale=0.8]{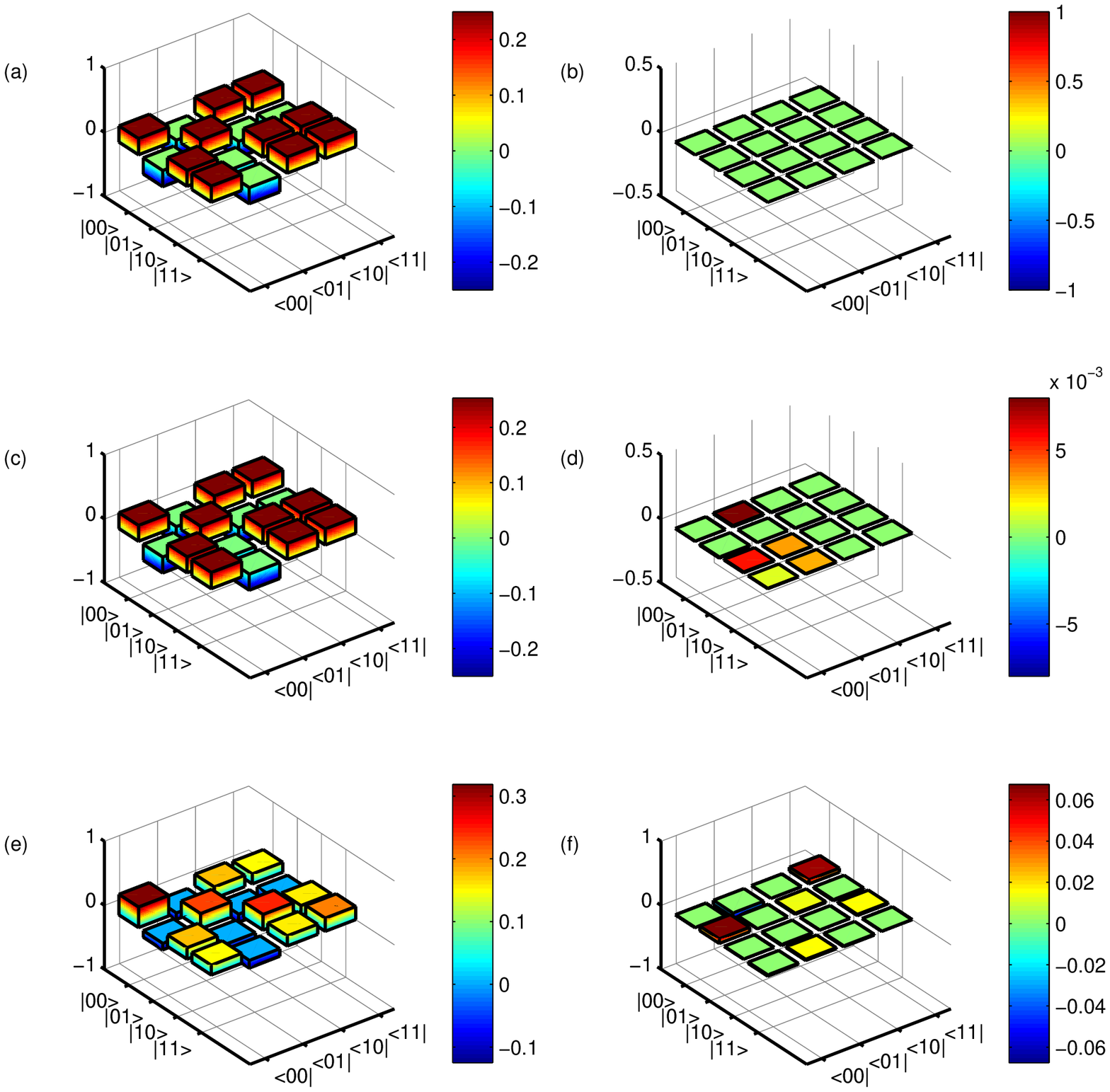}
\caption{LHS: Real part of the density matrix for the Z-state $\Ket{Z_2^1}$. RHS: Imaginary part of the density matrix for the Z-state $\Ket{Z_2^1}$. (a) and (b): ideal density matrix, (c) and (d): simulated density matrix, (e) and (f): experimentally obtained density matrix}

\label{Figure8}
\end{figure}

\begin{figure}[H]
\centering
\includegraphics[scale=0.4]{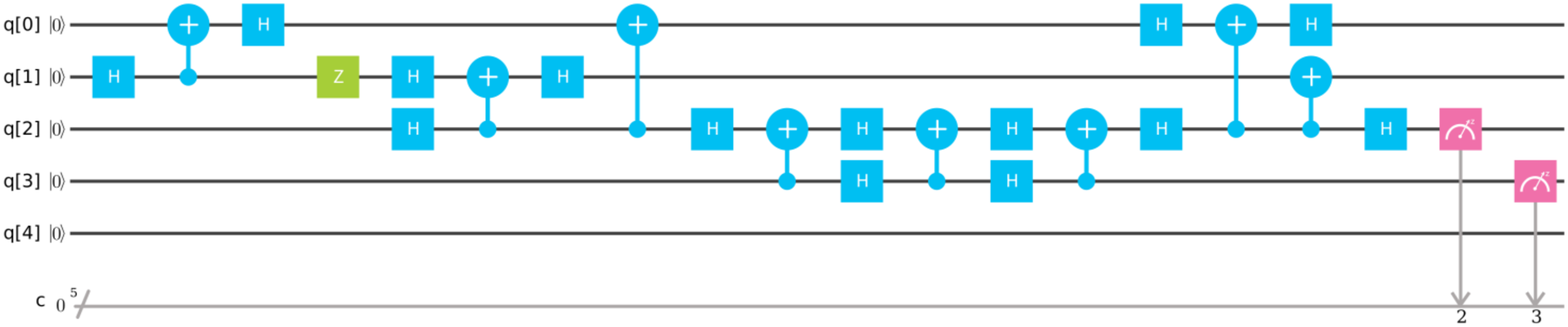}
\caption{Circuit designed in ibmqx4 to measure the 2 ancillary qubits}
\label{Figure9}
\end{figure}

\begin{figure}[H]
\centering
\includegraphics[scale=0.8]{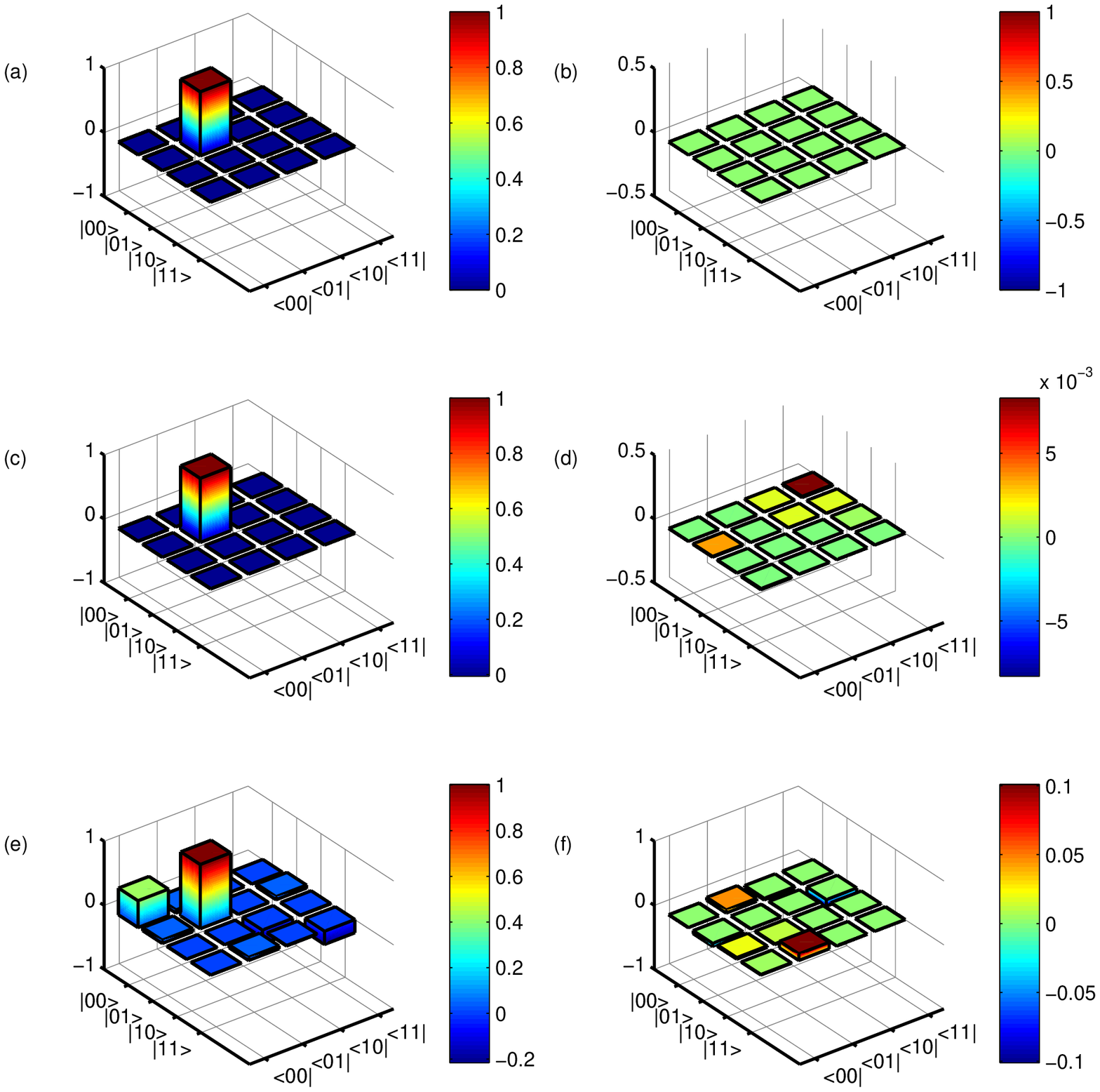}
\caption{LHS: Real part of the density matrix for the ancillary qubits corresponding to the Z-state $Z_2^1$. RHS: Imaginary part of the density matrix for the ancillary qubits corresponding to the Z-state $Z_2^1$. (a) and (b): ideal density matrix, (c) and (d): simulated density matrix, (e) and (f): experimentally obtained density matrix.}
\label{Figure10}
\end{figure}

We measure the ancilla qubits for obtaining the information about the Z-state. Fig. \ref{Figure9} illustrates the the quantum circuit for measuring the ancilla states. As the input Z-state is $\Ket{Z_2^1}$, from the Table 1, it is predicted that the ancilla state should be in $|01\rangle$ state. Hence, the theoretical density matrix for ancilla state ($\rho^{T}$) is given as, $\rho^{T}=|01\rangle\langle01|$.  
The experimental density matrices (simulational-$\rho_S^E$, and run-$\rho_R^E$) for the same are provided in the following. The comparison of density matrices is depicted in the Fig. \ref{Figure10}. 

\[\begin{split}
\rho_{S}^{E} =
\left[{\begin{array}{cccc}
                  0.0000&-0.0023&0.0013&-0.0058\\
                  -0.0023&1.0000&0.0023&0.0013\\
                  0.0013&0.0023&0.0000&0.0048\\
                  -0.0058&0.0013&0.0048&0.0000\\
\end{array} } \right] +
\end{split}
\]
\[
\begin{split}
 \\ i\left[ {\begin{array}{cccc}
                              0.0000&-0.0035&0.0015&0.0083\\
                              0.0035&0.0000&0.0013&0.0015\\
                              -0.0015&-0.0013&0.0000&0.0005\\
                              -0.0083&-0.0015&-0.0005&0.0000\\
\end{array} } \right]
\end{split}
\]

 \[
 \begin{split}
   \rho_{R}^{E}=
  \left[ {\begin{array}{cccc}
   0.4170&0.0460&-0.0003&-0.0025\\
   0.0460&1.0030&-0.0035&0.0432\\
   -0.0003&-0.0035&-0.2160&0.0050\\
   -0.0025&0.0432&0.0050&-0.2040\\
  \end{array} } \right]+
  \end{split}
 \]
 \[
 \begin{split}
  \\ i\left[ {\begin{array}{cccc}
   0.0000&0.0448&-0.0220&-0.0062\\
   -0.0448&0.0000&-0.0112&-0.1010\\
   0.0220&0.0112&0.0000&0.0012\\
   0.0062&0.1010&-0.0012&0.0000\\
  \end{array} } \right]
  \end{split}
 \]

\section{Discussion and conclusion}
To conclude, we have defined here a new type of highly entangled state named as Z-state. We have proposed a new quantum circuit for non-destructively discriminating Z-states. We run the quantum circuit in the IBM quantum computer and verify the results by quantum state tomography. It is found that we have experimentally prepared the nondestructive Z-states with a fidelity of 0.815. We hope, Z-states can be applied to the branch of measurement-based quantum computation. We also could find the application of nondestructive discrimination of orthogonal entangled states in distributed quantum computing in a quantum network.

{\bf Acknowledgements.}
B.K.B. is financially supported by DST Inspire Fellowship. SS and KS acknowledge the support of HBCSE and TIFR for conducting National Initiative on Undergraduate Sciences (NIUS) Physics camp. We are extremely grateful to IBM team and IBM QE project. The discussions and opinions developed in this paper are only those of the authors and do not reflect the opinions of IBM or IBM QE team.

\end{document}